\documentclass[conference,10pt,letterpaper]{IEEEtran}
\IEEEoverridecommandlockouts

\usepackage{cite}
\usepackage{amsmath,amssymb,amsfonts}
\usepackage{graphicx}
\usepackage{makecell}
\usepackage{textcomp}
\usepackage{xcolor}
\usepackage{array}
\usepackage{multirow}
\usepackage{caption}
\usepackage{subcaption}
\usepackage{algpseudocode}
\usepackage{algorithm}
\usepackage{fancyhdr}
\fancypagestyle{firststyle}{\fancyhf{}
  \fancyhead[L]{\small X. Liu, M. Ying, X. Wang, D. Shakya, H. Nikbakht, D. Abraham, and T. S. Rappaport, ``Map-Free Single-Anchor Position Localization Using Multipath Uncertainty at Upper Mid-Band," in \textit{2026 IEEE Global Communications Conference (GLOBECOM)}, Macao, China, Dec. 2026, pp. 1--6.}
 }

\newcolumntype{R}[1]{>{\raggedleft\arraybackslash}p{#1}}
\newcolumntype{C}[1]{>{\centering\arraybackslash}p{#1}}
\newcommand{\edit}[1]{#1}

\def\BibTeX{{\rm B\kern-.05em{\sc i\kern-.025em b}\kern-.08em
    T\kern-.1667em\lower.7ex\hbox{E}\kern-.125emX}}

\begin{document}

\title{Map-Free Single-Anchor Position Localization Using Multipath Uncertainty at Upper Mid-Band\thanks{This paper was supported by the NYU WIRELESS Industrial Affiliates Program, NYU Tandon ECE PhD Fellowship, and NSF Grant No. 2234123.}}

\author{\IEEEauthorblockN{
Xingchen Liu\IEEEauthorrefmark{1}, 
Mingjun Ying, 
Xinquan Wang, 
Dipankar Shakya,
Homa Nikbakht,\\
Daniel Abraham,
and Theodore S. Rappaport\IEEEauthorrefmark{2}}
\IEEEauthorblockA{NYU WIRELESS, New York University, Brooklyn, NY 11201, USA}
\IEEEauthorblockA{\{xl5933\IEEEauthorrefmark{1}, tsr\IEEEauthorrefmark{2}\}@nyu.edu}
}

\maketitle
\thispagestyle{firststyle}
\bstctlcite{BSTcontrol}
\begin{abstract}
Future 5G-Advanced and 6G systems will exploit directional wideband channels for path-based localization, yet map-free indoor positioning remains challenging because extracted multipath components (MPCs) vary widely in localization reliability. This paper presents an uncertainty-aware map-free single-anchor localization framework using 16.95~GHz indoor directional channel measurements collected at NYU WIRELESS in Brooklyn, New York. Per-MPC angular covariance is estimated from the local consistency of neighboring pattern-de-embedded power delay profile (PDP) observations and propagated into geometric localization through weighted fusion of point-type and line-type constraints. Measurements across 20 links with TX--RX separations from 11~m to 97~m show that the proposed covariance-weighted method achieves a mean localization error of 3.86~m, a median of 2.66~m, and 70\% of links within 5~m using five retained MPCs. It outperforms the power-weighted and unweighted variants under identical point/line constraints, as well as the literature baselines. These results indicate that local directional consistency provides reliability information beyond received power alone and demonstrate a practical map-free approach for future indoor directional systems.

\end{abstract}


\section{Introduction}

Future 5G-Advanced and 6G systems are expected to support high-data-rate communications, accurate user localization, and environmental awareness. In particular, upper-mid-band frequencies, often referred to as Frequency Range~3, together with millimeter-wave (mmWave) and sub-terahertz bands, provide large bandwidths and support highly directional transmission and reception, enabling finer separation of multipath components (MPCs) in delay and angle than conventional microwave systems~\cite{Nie2013PIMRC,Sun2014ICC,9500385,490227, 7414164}. This improved delay and angular resolution enables multipath-based positioning, in which the estimated angle of departure (AOD), angle of arrival (AOA), and time of flight (ToF) of MPCs between a transmitter (TX) and a receiver (RX) are converted into geometric constraints on the RX position~\cite{Shahmansoori2018TWC,Kanhere2025TWC,rappaport2004raytracing}.

Multipath-based localization has been investigated from several perspectives. Early single-anchor formulations related AOD, AOA, and ToF measurements of non-line-of-sight (NLOS) paths to the RX position through explicit single-bounce reflection geometry~\cite{8445980}, while a related 3D zero-forcing joint-positioning (ZF--JP) method jointly estimated the RX and scatterer positions~\cite{Wei2011PIMRC}. In mmWave multiple-input multiple-output systems, single-transmitter localization has been analyzed using delay and angular information from line-of-sight (LOS) and NLOS paths, together with performance bounds and practical estimators~\cite{Shahmansoori2018TWC}. Reflected MPCs have also been represented by virtual anchors derived from known environmental geometry~\cite{witrisal2016high}. More recent single-base-station methods jointly exploit LOS and NLOS MPCs for snapshot-based position and orientation estimation~\cite{Nazari2023TVT}, whereas map-assisted techniques use angular and timing information with environmental priors to account for higher-order propagation~\cite{Kanhere2025TWC}. Measured power delay profiles have also been used directly for position estimation, where a candidate location is refined by aligning ray-traced and measured PDPs~\cite{ying2025multistage}.

Despite this progress, many map-free methods insert estimated path parameters directly into ideal LOS or single-bounce geometric models~\cite{8445980,Wei2011PIMRC,Shahmansoori2018TWC,Nazari2023TVT}. Map-assisted methods can validate or associate individual paths using environmental priors~\cite{Kanhere2025TWC,witrisal2016high}, but accurate maps may be unavailable, incomplete, or costly to acquire and maintain. Without such priors, estimated MPCs can differ substantially when acting as geometric constraints because of multi-bounce propagation, diffuse scattering, and path mixing. Recent posterior-based formulations report a distribution over candidate locations rather than a single point estimate~\cite{lei2026beyond}, and a ray-tracing digital twin can be used to score each candidate against the measured multipath profile~\cite{lei2026locus}. These limitations motivate a measurement-driven estimate of path-specific reliability from local directional observations, together with explicit propagation of the resulting uncertainty into the position estimate.

To address this gap, this paper presents a map-free, measurement-driven single-anchor localization framework that estimates path-specific reliability directly from indoor directional measurements. The main contributions are as follows. \edit{First, building on the antenna-de-embedded multipath extraction (ADME) procedure~\cite{ju2024statistical}, we use neighboring directional power delay profile (PDP) observations to refine MPC parameters and estimate a per-MPC angular covariance from local directional consistency. Second, we propagate the estimated covariance through the geometric construction and develop a covariance-weighted solver that fuses LOS and feasible single-bounce point constraints with line fallbacks. Finally, experiments on 20 measured indoor links demonstrate that the proposed method outperforms power-weighted and unweighted variants constructed from the same retained MPCs and geometric constraints, as well as the planar NLOS method in~\cite{8445980} and the 3D ZF--JP method in~\cite{Wei2011PIMRC}}.

\section{MPC Estimation and Uncertainty Modeling}
\label{sec:3}

\subsection{Indoor Hotspot (InH) Measurement Campaign Overview}
\label{sec:meas}

The 16.95~GHz InH measurements used in this paper were conducted at the NYU WIRELESS Research Center in Brooklyn, NY \cite{shakya2024comprehensive, shakya2025angular}, using a sliding-correlation channel sounder. \edit{The campaign spans enclosed rooms, open work areas, and connecting corridors and contains 7 LOS and 13 NLOS links with TX--RX separations from 11~m to 97~m.} For each TX--RX pair, directional PDPs are collected over discrete TX/RX steering directions in both azimuth and elevation, forming a 3D directional dataset \cite{rappaport2004building}, enabling finer resolution of MPC directions and per-path uncertainty quantification.


\subsection{MPC extraction and refinement overview}
\label{subsec:mpc_extraction}

Directional PDPs over discrete TX/RX steering pairs are first aggregated into coarse dominant MPCs with initial delay and AOD/AOA estimates~\cite{samimi20163}. Each candidate is then refined using ADME~\cite{ju2024statistical}, which collects observations from steering pairs within approximately one half-power beamwidth of the coarse direction. For the $l$-th MPC with delay $\hat{\tau}_l$, the neighborhood power is computed as
\begin{equation}
p^{(l)}_{i,j}\triangleq
\sum_{n\in\mathcal{W}(\hat{\tau}_l)}
\mathrm{PDP}_{i,j}[n],
\qquad (i,j)\in\mathcal{N}_l,
\label{eq:p_win_def}
\end{equation}
where $\mathcal{W}(\hat{\tau}_l)$ is the delay window centered at $\hat{\tau}_l$, and $\mathcal{N}_l$ is the directional neighborhood. ADME searches nearby candidates and selects the direction yielding the most consistent pattern-de-embedded neighborhood powers. The refined parameters can be interpreted as minimizing the sum of squared neighborhood residuals over the local angular search region.

\subsection{Local angular uncertainty modeling}
\label{subsec:mpc_uncertainty}

For the $l$-th MPC, let $\hat{\boldsymbol{\theta}}_l
=
\left[
\hat{\varphi}^{\rm az}_{t,l},
\hat{\varphi}^{\rm el}_{t,l},
\hat{\varphi}^{\rm az}_{r,l},
\hat{\varphi}^{\rm el}_{r,l}
\right]^\top$
denote its angle estimate, and let $\hat P_l$ denote the corresponding power. Because different MPCs may be supported by varying numbers of consistent neighborhood observations, the angular estimates do not share a common level of reliability. To capture this path-dependent directional uncertainty, we seek a local angular covariance matrix $\mathbf{R}^{(\mathrm{ang})}_l\in\mathbb{R}^{4\times 4}$ for each MPC, estimated from the neighborhood observations introduced in Section~\ref{subsec:mpc_extraction}.

\subsubsection{Local residual model and weighting}

The proposed uncertainty model is built from the local consistency of the neighborhood observations in \eqref{eq:p_win_def}. For the $l$-th MPC, the window-integrated power measured at each neighboring steering pair $(i,j)\in\mathcal{N}_l$ is modeled as
\begin{equation}
p^{(l)}_{i,j}
\approx
P_l\, g_{i,j}(\boldsymbol{\theta}_l)+\epsilon^{(l)}_{i,j},
\label{eq:power_model_4d}
\end{equation}
where $P_l$ is the de-embedded MPC power, $\epsilon^{(l)}_{i,j}$ captures residual disturbance, and $g_{i,j}(\boldsymbol{\theta}_l)$ is the combined TX/RX horn power gain evaluated at the angular offset between the steering direction and the MPC direction.

Let $\boldsymbol{\eta}_l=[\boldsymbol{\theta}_l^\top,P_l]^\top\in\mathbb{R}^{5}$ denote the local parameter vector. For each neighborhood observation $(i,j)\in\mathcal{N}_l$, define the residual function
\begin{equation}
r^{(l)}_{i,j}(\boldsymbol{\eta}_l)
\triangleq
\frac{p^{(l)}_{i,j}}{g_{i,j}(\boldsymbol{\theta}_l)}-P_l.
\label{eq:rij_def}
\end{equation}
The residuals are then evaluated at the estimated $\hat{\boldsymbol{\eta}}_l$ to measure how tightly the neighboring observations cluster around $\hat P_l$. Larger residual indicates weaker local support, which may arise from low signal-to-noise ratio (SNR), pattern mismatch, or MPC mixing within the delay window. 

Then, the effective local residual scale based on \cite{bates1988nonlinear} is estimated by
\begin{equation}
\widehat{\sigma}_l^2
=
\frac{\sum_{(i,j)\in\mathcal{N}_l}   (r^{(l)}_{i,j})^{\,2}}{|\mathcal{N}_{l}|-5},
\label{eq:sigmahat}
\end{equation}
where the denominator reflects the residual degrees of freedom for the five parameters. The estimated residual scale $\widehat{\sigma}_l^2$ absorbs not only thermal noise but also non-ideal disturbances that are difficult to separate explicitly in measured directional PDPs, and should therefore be interpreted as a local uncertainty scale rather than a purely thermal-noise variance.

\subsubsection{Covariance approximation from neighborhood observations}
\label{covariance}
We approximate the local covariance of $\hat{\boldsymbol{\eta}}_{l}$ by locally linearizing the nonlinear least-squares residual model, following standard regression results~\cite{bates1988nonlinear}. Let $\mathbf{r}_{l}(\boldsymbol{\eta}_{l})\in\mathbb{R}^{M_{l}}$ stack $r^{(l)}_{i,j}(\boldsymbol{\eta}_{l})$ over $(i,j)\in\mathcal{N}_{l}$, and define its Jacobian as 
\begin{equation}
\mathbf{J}_{l}\triangleq\left.\frac{\partial\mathbf{r}_{l}(\boldsymbol{\eta}_{l})}{\partial\boldsymbol{\eta}_{l}^{\top}}\right|_{\boldsymbol{\eta}_{l}=\hat{\boldsymbol{\eta}}_{l}}\in\mathbb{R}^{M_{l}\times 5}.     
\end{equation}
Its four angular columns are evaluated from the interpolated antenna-pattern tables using finite differences and therefore reflect the local pattern slopes, whereas the power derivative is obtained analytically as $\partial r^{(l)}_{i,j}(\boldsymbol{\eta}_{l})/\partial P_{l}=-1$.

\edit{Under this local linearization, perturbations in the stacked residual vector induce corresponding perturbations in the estimated parameters through the least-squares solution. Treating the neighborhood residuals as locally uncorrelated with a common variance $\widehat{\sigma}_{l}^{2}$, standard covariance propagation gives}
\begin{equation}
\mathrm{Cov}(\hat{\boldsymbol{\eta}}_l)
\approx
\widehat{\sigma}_l^2
\left(\mathbf{J}_l^\top \mathbf{J}_l\right)^{-1}.
\label{eq:cov_eta}
\end{equation}
The right side of \eqref{eq:cov_eta} is used as a practical local covariance approximation. Here, $\widehat{\sigma}_{l}^{2}$ captures the effective residual mismatch, whereas $(\mathbf{J}_{l}^{\top}\mathbf{J}_{l})^{-1}$ characterizes the local parameter identifiability. Since $\hat{\boldsymbol{\eta}}_l=[\hat{\boldsymbol{\theta}}_l^\top,\hat P_l]^\top$ contains the four angle parameters and one power parameter, only the angular block is retained for later localization. The desired angular covariance is therefore taken as the $4\times4$ principal submatrix
\begin{equation}
\mathbf{R}^{(\mathrm{ang})}_l
=
\left[\mathrm{Cov}(\hat{\boldsymbol{\eta}}_l)\right]_{1:4,1:4}.
\label{eq:cov_ang}
\end{equation}

\begin{figure}[t]
  \centering
  \begin{subfigure}[t]{\columnwidth}
    \centering
    \includegraphics[width=0.9\columnwidth]{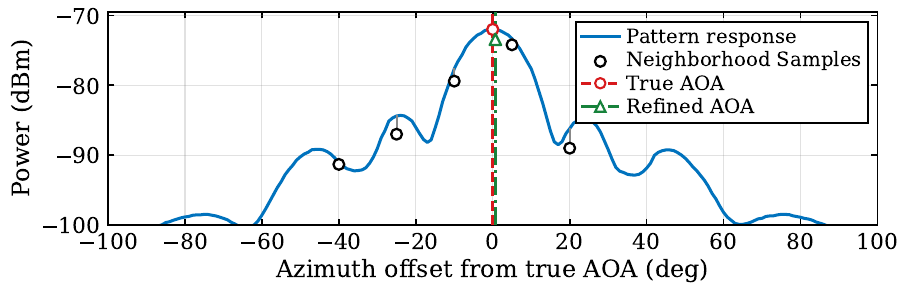}
    \caption{\edit{Well-supported neighborhood with an accurate refined AOA.}}
    \label{fig:obs_good}
  \end{subfigure}

  \vspace{1pt}

  \begin{subfigure}[t]{\columnwidth}
    \centering
    \includegraphics[width=0.9\columnwidth]{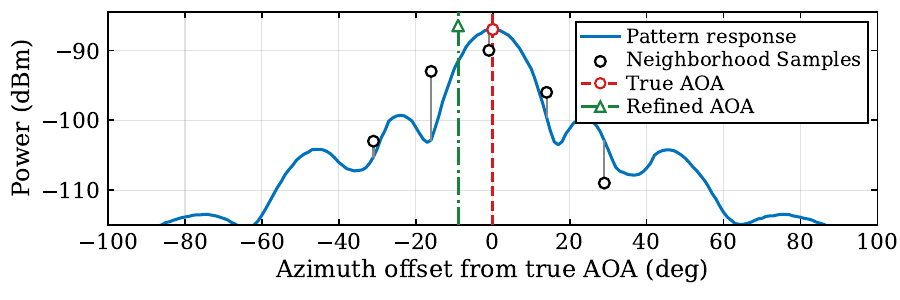}
    \caption{\edit{Poorly supported neighborhood with larger pattern mismatch and a biased refined AOA.}}
    \label{fig:obs_bad}
  \end{subfigure}

  \caption{\edit{Examples of neighborhood support on a fixed-elevation slice. The blue curve, black circles, and red/green lines denote the antenna pattern, measured samples, and true/refined AOAs, respectively; gray segments indicate sample--pattern mismatch.}}
  \label{fig:obs_good_bad}
\end{figure}

\edit{The covariance approximation is used only when it is numerically well defined. Specifically, MPCs with $M_l\leq5$ are excluded because the residual degrees of freedom are insufficient, while MPCs with nearly singular $\mathbf{J}_l^\top\mathbf{J}_l$ are excluded because their angular covariance cannot be reliably estimated.}

\edit{For the retained MPCs, Fig.~\ref{fig:obs_good_bad} illustrates the effect of local directional consistency. In Fig.~\ref{fig:obs_good}, close agreement with the antenna pattern yields tightly clustered de-embedded powers and a small angular covariance. In Fig.~\ref{fig:obs_bad}, larger mismatch increases the residual spread, may bias the refined AOA, and leads to a larger covariance. The samples shown are one-dimensional slices of the full four-dimensional neighborhoods.}

\edit{The PTP-based synchronization and periodic reference-PDP realignment achieve sub-nanosecond absolute timing accuracy~\cite{shakya2023sub}, corresponding to a synchronization-induced one-way path-length uncertainty below 0.3~m and well below the observed localization errors. We therefore treat $d_{l}=c\widehat{\tau}_{l}$ as fixed in the subsequent uncertainty propagation; delay errors caused by noise or mixed MPCs remain for future joint angle--delay uncertainty modeling.}

\section{Single-Anchor Localization}
\label{sec:loc}

\subsection{Problem formulation}
\label{subsec:loc_setup}

Consider single-anchor localization with a known TX position and an unknown RX position, where the TX and RX heights are fixed at $h_t$ and $h_r$, respectively. Let the known TX position be
$\mathbf{p}_t=[x_t,y_t,h_t]^\top$, and let the unknown horizontal RX position be
$\mathbf{x}=[x,y]^\top$. The corresponding 3D RX position is $\mathbf{p}_r=[x,y,h_r]^\top=[\mathbf{x}^\top,h_r]^\top$.
Our goal is to estimate $\mathbf{x}$ from the refined MPC delays and angles with uncertainty. No environment prior is assumed in our setting: for NLOS paths, the reflection/scatter point is unknown and is not associated with any map or surface model.

\subsection{MPC gating}
\label{subsec:loc_gating}

Before constructing localization constraints, the available MPCs are sorted by delay, and only the first $K$ arrivals are retained to favor LOS and low-order reflected paths, since later arrivals are more likely to arise from higher-order reflections, MPC mixing, or diffuse scattering. Because delay alone cannot reliably distinguish single-bounce from higher-order paths, treating the latter with a single-bounce model can cause large localization errors~\cite{Kanhere2025TWC}. Covariance weighting therefore acts as a secondary soft gate: MPCs with inconsistent de-embedded neighborhood powers receive larger angular covariances and consequently less influence on the position estimate.

\subsection{MPC-induced geometric constraints}
\label{subsec:loc_constraints}

\edit{For the $l$-th retained MPC, let $d_l=c\hat{\tau}_l$ denote the path length obtained from the calibrated delay, whose residual uncertainty is neglected as discussed in Section~\ref{covariance}.} The refined departure and arrival angles define the unit direction vectors
\begin{equation}
\mathbf{u}_{t,l}=\mathbf{u}\!\left(\hat{\varphi}^{\rm az}_{t,l},\hat{\varphi}^{\rm el}_{t,l}\right),
\qquad
\mathbf{u}_{r,l}=\mathbf{u}\!\left(\hat{\varphi}^{\rm az}_{r,l},\hat{\varphi}^{\rm el}_{r,l}\right),
\label{eq:u_vec_def}
\end{equation}
where $\mathbf{u}_{t,l}$ points from the TX along the departure direction, and $\mathbf{u}_{r,l}$ points from the RX toward the source of the arriving path. We further denote the horizontal projections by $\mathbf{u}^{\rm xy}_{t,l}$ and $\mathbf{u}^{\rm xy}_{r,l}$, the vertical components by $u^{(z)}_{t,l}$ and $u^{(z)}_{r,l}$, and the TX horizontal position by $\mathbf{p}_t^{\rm xy}=[x_t,y_t]^\top$.

\subsubsection{LOS point constraint}
Let $l_0\triangleq\arg\min_l \hat{\tau}_l$ denote the earliest-arriving retained MPC. We test the LOS hypothesis only on this MPC, since later paths are unlikely to be LOS. An LOS hypothesis is declared if the departure and arrival directions are approximately reciprocal:
\begin{equation}
\left\|\mathbf{u}_{t,l_0}+\mathbf{u}_{r,l_0}\right\|\le \epsilon_{\rm los}.
\label{eq:los_gate}
\end{equation}
For a LOS path, either the departure or the arrival direction alone, combined with the path delay, is sufficient to determine the RX position. Fusing both directions allows the more reliably estimated side to dominate.
Specifically, let $\sigma_{t,l_0}^2$ and $\sigma_{r,l_0}^2$ denote the traces of AOD- and AOA-related covariance sub-blocks, i.e., the sums of the corresponding azimuth and elevation variances. We use these traces as compact measures of overall angular uncertainty, so that the less certain side contributes less to the fused LOS direction. We then define
\begin{equation}
\tilde{\mathbf{u}}_{l_0}
=
\frac{
w_{t,l_0}\mathbf{u}_{t,l_0}
-
w_{r,l_0}\mathbf{u}_{r,l_0}
}{
\left\|
w_{t,l_0}\mathbf{u}_{t,l_0}
-
w_{r,l_0}\mathbf{u}_{r,l_0}
\right\|
},
\label{eq:los_fused_dir}
\end{equation}
where $w_{t,l_0}=1/\sigma_{t,l_0}^2$, $w_{r,l_0}=1/\sigma_{r,l_0}^2$. 
The resulting 3D RX candidate is constructed as
\begin{equation}
\mathbf{p}^{(p)}_{r,l_0}=\mathbf{p}_t+d_{l_0}\tilde{\mathbf{u}}_{l_0}.
\end{equation}
We keep this candidate only if its height is consistent with the RX height, i.e.,
$\left|(\mathbf{p}^{(p)}_{r,l_0})_z-h_r\right|\le\epsilon_z$.
Its horizontal projection,
\[
\mathbf{y}_{l_0} \triangleq (\mathbf{p}^{(p)}_{r,l_0})_{1:2},
\]
is then used as a point-type geometric constraint for the unknown horizontal RX position $\mathbf{x}$. Accordingly, the associated point residual is defined as the 2D vector
\[
\mathbf{r}^{(p)}_{l_0}(\mathbf{x})=\mathbf{x}-\mathbf{y}_{l_0}.
\]

\subsubsection{Single-bounce NLOS point constraint and line fallback}
For an NLOS MPC, we first test a single-bounce hypothesis with an unknown reflection/scatter point $\mathbf{s}_l$. Under the direction convention above, the geometry satisfies
$\mathbf{s}_l=\mathbf{p}_t+t_l\mathbf{u}_{t,l}$,
$\mathbf{p}_r=\mathbf{s}_l-r_l\mathbf{u}_{r,l}$,
and $t_l+r_l=d_l$, where $t_l$ and $r_l$ are the TX-to-scatterer and scatterer-to-RX segment lengths, respectively. Enforcing the fixed RX height gives
\begin{equation}
t_l=\frac{(h_r-h_t)+d_l u^{(z)}_{r,l}}{u^{(z)}_{t,l}+u^{(z)}_{r,l}},
\qquad
r_l=d_l-t_l.
\label{eq:sb_t_r}
\end{equation}
If this solution is feasible, i.e., $t_l\ge 0$, $r_l\ge 0$, and the denominator in \eqref{eq:sb_t_r} is well-conditioned, we obtain the horizontal RX point candidate
\begin{equation}
\mathbf{y}_{l}
=
\mathbf{p}_t^{\rm xy}+t_l\mathbf{u}^{\rm xy}_{t,l}-r_l\mathbf{u}^{\rm xy}_{r,l},
\label{eq:nlos_point}
\end{equation}
which induces the same point residual form, namely $\mathbf{r}^{(p)}_{l}(\mathbf{x})=\mathbf{x}-\mathbf{y}_{l}$.

\begin{figure}[t]
    \centering
    \includegraphics[width=0.65\columnwidth]{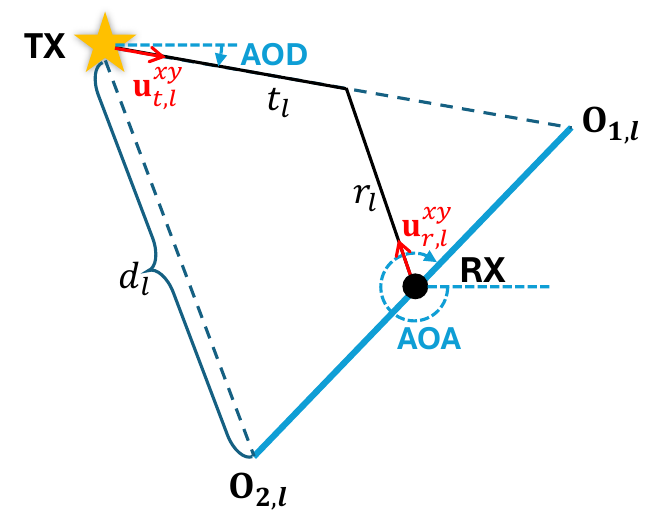}
    \caption{Illustration of the NLOS line constraint on the horizontal plane. The two endpoint hypotheses $\mathbf{O}_{1,l}$ and $\mathbf{O}_{2,l}$ are determined by the AOD, AOA, and total path length $d_l$. Any horizontal RX location consistent with a valid split $t_l+r_l=d_l$ lies on the line joining these two points.}
    \label{fig:line_constraint}
\end{figure}

When the single-bounce point construction is infeasible or numerically unstable, a line-type constraint is used instead based on the horizontal single-bounce geometry. Specifically, using the total path length $d_l$, the TX departure direction $\mathbf{u}^{\rm xy}_{t,l}$, and the RX arrival direction $\mathbf{u}^{\rm xy}_{r,l}$, two endpoint hypotheses are defined on the horizontal plane as
\begin{equation}
\mathbf{O}_{1,l}=\mathbf{p}_t^{\rm xy}+d_l\,\mathbf{u}^{\rm xy}_{t,l},
\qquad
\mathbf{O}_{2,l}=\mathbf{p}_t^{\rm xy}-d_l\,\mathbf{u}^{\rm xy}_{r,l},
\label{eq:line_endpoints}
\end{equation}
as illustrated in Fig.~\ref{fig:line_constraint}. Any horizontal RX location consistent with a valid split $t_l+r_l=d_l$ would lie on the line joining $\mathbf{O}_{1,l}$ and $\mathbf{O}_{2,l}$, i.e.,
\begin{equation}
\mathcal{L}_l=\left\{\mathbf{O}_{1,l}+\lambda\left(\mathbf{O}_{2,l}-\mathbf{O}_{1,l}\right):\lambda\in\mathbb{R}\right\}.
\label{eq:line_set}
\end{equation}
The associated line residual $r^{(\ell)}_l(\mathbf{x})$ is then defined as the signed perpendicular distance from the candidate RX position $\mathbf{x}$ to $\mathcal{L}_l$.

\subsection{Weighted position estimation}
\label{subsec:loc_weighting}

Each retained MPC contributes either a point-type or a line-type constraint on $\mathbf{x}$, producing a 2D vector residual or a scalar residual, respectively. We incorporate both in a unified type-aware weighted formulation.

Let $\mathcal{P}$ denote the retained MPCs yielding valid point constraints (LOS or feasible single-bounce), and let $\mathcal{L}$ denote those yielding line constraints. For each $l\in\mathcal{P}$, the point candidate $\mathbf{y}_l$ depends on the estimated angular parameters $\boldsymbol{\theta}_l$, and its residual is
\begin{equation}
\mathbf{r}^{(p)}_l(\mathbf{x})
=
\mathbf{x}-\mathbf{y}_l.
\end{equation}
Its uncertainty is obtained by first-order propagation of the angular covariance $\mathbf{R}^{(\mathrm{ang})}_l$ through the point construction:
\begin{equation}
\mathbf{\Sigma}^{(p)}_l
\approx
\mathbf{G}^{(p)}_{\theta,l}\,
\mathbf{R}^{(\mathrm{ang})}_l\,
\mathbf{G}^{(p)\top}_{\theta,l},
\qquad
\mathbf{G}^{(p)}_{\theta,l}
\triangleq
\left.
\frac{\partial \mathbf{y}_l}
{\partial \boldsymbol{\theta}_l}
\right|_{\hat{\boldsymbol{\theta}}_l}.
\label{eq:point_cov_prop}
\end{equation}
Since $\mathbf{x}$ is deterministic during residual evaluation, $\mathbf{\Sigma}^{(p)}_l$ is also the covariance of $\mathbf{r}^{(p)}_l(\mathbf{x})$.

For each $l\in\mathcal{L}$, $r^{(\ell)}_l(\mathbf{x})$ is the signed perpendicular distance from $\mathbf{x}$ to the induced line. Its variance depends on $\mathbf{x}$ because angular perturbations rotate the line, and is therefore approximated at a pilot estimate $\mathbf{x}^{(0)}$ obtained from the unweighted solution over the same constraint set:
\begin{equation}
(\bar{\sigma}^{\ell}_l)^2
\approx
\mathbf{g}^{(\ell)}_{\theta,l}(\mathbf{x}^{(0)})
\mathbf{R}^{(\mathrm{ang})}_l
\mathbf{g}^{(\ell)\top}_{\theta,l}(\mathbf{x}^{(0)}),
\
\mathbf{g}^{(\ell)}_{\theta,l}(\mathbf{x})
\triangleq
\left.
\frac{\partial r^{(\ell)}_l(\mathbf{x})}
{\partial \boldsymbol{\theta}_l}
\right|_{\hat{\boldsymbol{\theta}}_l}.
\label{eq:line_var_prop}
\end{equation}

The final RX position estimate is obtained from
\begin{equation}
\begin{aligned}
\hat{\mathbf{x}}
=
\arg\min_{\mathbf{x}\in\mathbb{R}^2}
\left(
\sum_{l\in\mathcal{P}}
\mathbf{r}^{(p)\top}_l(\mathbf{x})
\big(\mathbf{\Sigma}^{(p)}_l+\epsilon\mathbf{I}\big)^{-1}
\mathbf{r}^{(p)}_l(\mathbf{x})
\right. \\
\qquad\qquad\qquad\qquad\qquad
\left.
+\sum_{l\in\mathcal{L}}
\frac{\big(r^{(\ell)}_l(\mathbf{x})\big)^2}
{(\bar{\sigma}^{\ell}_l)^2+\epsilon}
\right),
\end{aligned}
\label{eq:joint_wls}
\end{equation}
where $\epsilon>0$ prevents numerical instability when a propagated covariance or variance is nearly singular. With these quantities fixed, \eqref{eq:joint_wls} is quadratic in $\mathbf{x}$ and admits a closed-form solution.

\edit{Algorithm~\ref{alg:framework} summarizes the complete processing chain developed in Sections~II and III, from directional MPC extraction and uncertainty estimation to geometric constraint construction and covariance-weighted position fusion.}

\begin{algorithm}[H]
\caption{Uncertainty-aware single-anchor localization}
\label{alg:framework}
\scriptsize
\begin{algorithmic}[1]
\Require Directional PDPs, known TX position and heights, retained MPC number $K$
\Ensure Estimated horizontal RX position $\hat{\mathbf{x}}$
\State Extract and refine MPC parameters using ADME
\State Estimate $\mathbf{R}^{(\mathrm{ang})}_l$; discard unsupported or ill-conditioned MPCs
\State Retain the $K$ earliest remaining MPCs and denote the earliest one by $l_0$
\For{each retained MPC $l$}
    \If{$l=l_0$ and the LOS reciprocity and height tests hold}
        \State Construct an LOS point constraint
    \ElsIf{the single-bounce point construction is feasible}
        \State Construct a single-bounce point constraint
    \Else
        \State Construct a line constraint
    \EndIf
\EndFor
\State Propagate angular uncertainty to the point covariances
\State Obtain $\mathbf{x}^{(0)}$ and compute the line-residual variances if line constraints exist
\State Solve~\eqref{eq:joint_wls} and \Return $\hat{\mathbf{x}}$
\end{algorithmic}
\end{algorithm}

\section{Numerical Results}
\label{sec:results}

The experiments are conducted on the 16.95~GHz indoor directional measurement dataset described in Section~\ref{sec:meas}, which contains 7 LOS and 13 NLOS links. The TX and RX heights are fixed at 2.4~m and 1.5~m, respectively. Localization error is measured as the 2D Euclidean distance between the estimated and ground-truth RX positions.

We evaluate three internal variants and two literature baselines. The three internal variants use the same retained MPCs and the same point/line constraint construction described in Section~\ref{subsec:loc_constraints}, but differ in their reliability weighting. The proposed covariance-weighted method (CW) uses the propagated angular uncertainty, the power-weighted method (PW) uses the de-embedded MPC power, and the unweighted method (UW) assigns equal weight to every constraint. This comparison therefore isolates the effect of measurement-derived covariance weighting from power-based and uniform weighting.

\edit{For external comparison, we consider two classical MPC-based localization methods. The planar NLOS geometric estimator in~\cite{8445980}, referred to hereafter as Planar NLOS, uses the azimuth AOD, azimuth AOA, and ToF of the retained MPCs; elevation information is omitted to remain consistent with its original 2D formulation. The 3D ZF--JP algorithm in~\cite{Wei2011PIMRC} jointly estimates the RX and scatterer positions using azimuth and elevation AOD/AOA measurements together with ToF. Its estimated horizontal RX coordinates are used to compute the 2D localization error. Both literature baselines use the same retained MPC set but follow their original single-bounce NLOS models, without the proposed LOS reciprocity test, point/line fallback, or measurement-derived reliability weighting.} For CW, PW, and UW, the LOS reciprocity threshold is set to $\epsilon_{\rm los}=0.087$, corresponding approximately to a $5^\circ$ mismatch from ideal reciprocity, and the height-consistency tolerance for point constraints is set to 0.5~m.

\begin{figure}[t]
    \centering
    \includegraphics[width=0.5\textwidth]{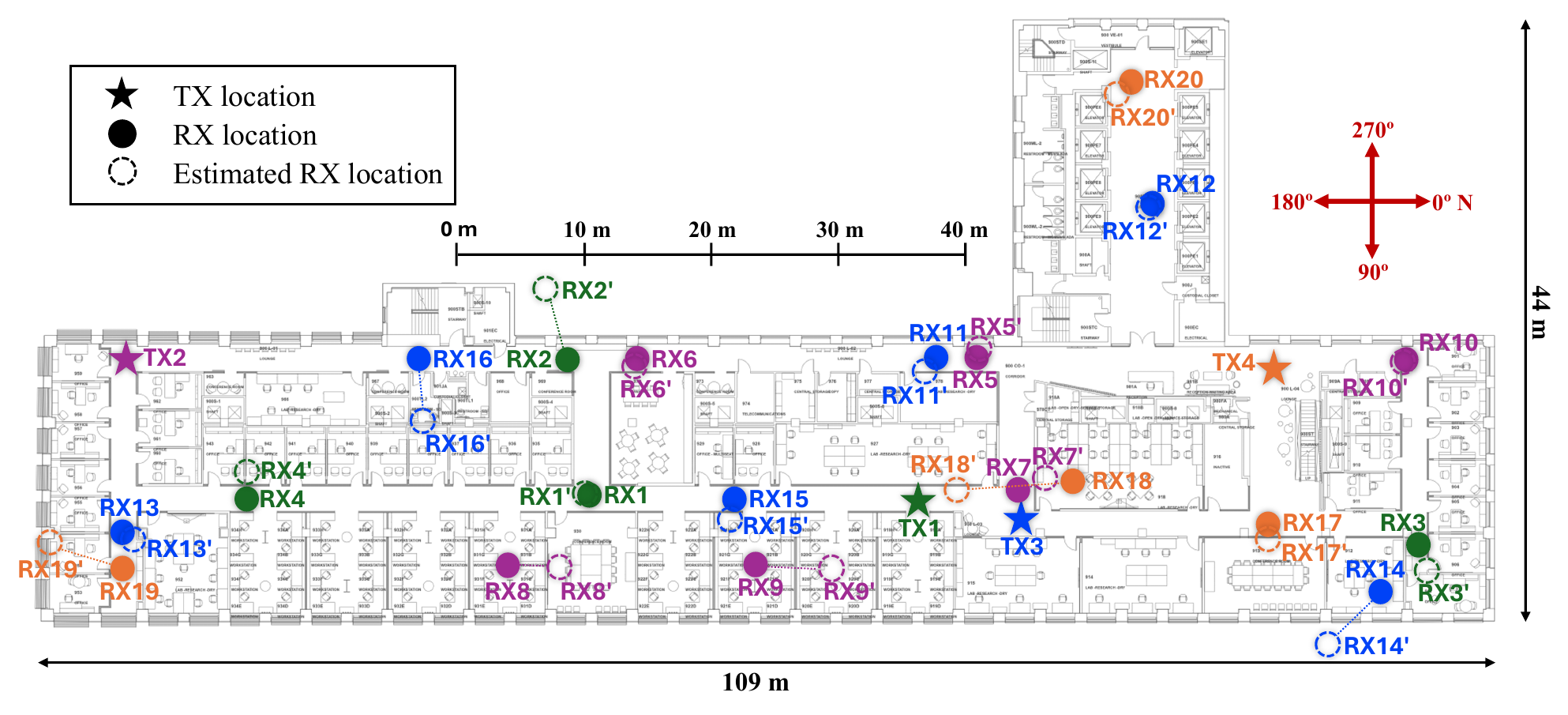}
    \caption{Localization results of the proposed method when $K=5$. Ground-truth and estimated RX locations are connected by dashed lines, with colors indicating the associated TX.
    }
    \label{fig:wide}
\end{figure}

\begin{table}[t]
\centering
\caption{Localization error comparison of three internal variants and two literature baselines using $K=5$ retained MPCs. Results are reported separately for LOS, NLOS, and all links.}
\label{tab:loc_compare_main}
\footnotesize
\setlength{\tabcolsep}{5.5pt}
\renewcommand{\arraystretch}{1.10}
\begin{tabular}{|c|c|c|c|c|}
\hline
\textbf{Method} & \textbf{Env.} & \textbf{Mean (m)} & \textbf{Median (m)} & \textbf{$<5$ m (\%)} \\
\hline
\multirow{3}{*}{\shortstack{\textbf{Covariance}\\\textbf{weighted}}}
& LOS & 1.86 & 1.41 & 100 \\
& NLOS & 4.94 & 4.57 & 53.84 \\
& \textbf{Overall} & \textbf{3.86} & \textbf{2.66} & \textbf{70} \\
\hline
\multirow{3}{*}{\shortstack{\textbf{Power}\\\textbf{weighted}}}
& LOS & 1.95 & 1.55 & 100 \\
& NLOS & 5.78 & 5.22 & 46.15 \\
& \textbf{Overall} & \textbf{4.44} & \textbf{4.01} & \textbf{65} \\
\hline
\multirow{3}{*}{\textbf{Unweighted}}
& LOS & 2.71 & 2.39 & 85.71 \\
& NLOS & 6.50 & 6.82 & 46.15 \\
& \textbf{Overall} & \textbf{5.17} & \textbf{4.35} & \textbf{60} \\
\hline
\multirow{3}{*}{\shortstack{\textbf{Planar }\\\textbf{NLOS}~\cite{8445980}}}
& LOS     & 4.27 & 4.15 & 71.43 \\
& NLOS    & 21.31 & 15.23 & 7.69 \\
& \textbf{Overall} & \textbf{15.35} & \textbf{7.46} & \textbf{30} \\
\hline
\multirow{3}{*}{\textbf{3D ZF--JP}~\cite{Wei2011PIMRC}}
& LOS & 3.89 & 3.29 & 71.43 \\
& NLOS & 15.22 & 9.87 & 15.38 \\
& \textbf{Overall} & \textbf{11.25} & \textbf{6.21} & \textbf{35} \\
\hline
\end{tabular}
\vspace{2pt}
\raggedright
\scriptsize
\\
The three internal methods use the same point/line constraint construction.
\end{table}


\subsection{Localization results of the proposed method}

Fig.~\ref{fig:wide} illustrates the localization results of the proposed method across all TX--RX pairs on the floor plan. LOS links are generally estimated with small errors, while larger errors occur for RXs deep inside rooms, such as TX3--RX14 and TX4--RX18, or with long TX--RX separations, like TX2--RX9 and TX4--RX19. These cases likely have limited single-bounce support, so the retained MPCs may induce point or line constraints that are inconsistent with the true RX location. Although covariance-based weighting reduces the influence of weakly supported paths, it cannot fully eliminate errors caused by geometric model mismatch. Overall, most RX positions are well estimated, with significant errors confined to a few challenging NLOS links.

\subsection{Localization performance comparison}
\label{subsec:compare_ablation}

Table~\ref{tab:loc_compare_main} presents the LOS, NLOS, and overall localization performance of the five methods. All methods perform better on LOS links, whereas NLOS links remain more challenging because of higher-order reflections, diffuse scattering, and MPC mixing. CW achieves the best overall performance, with a mean error of 3.86~m, a median error of 2.66~m, and 70\% of the links localized within 5~m. Compared with UW, CW reduces the overall mean and median errors from 5.17~m and 4.35~m to 3.86~m and 2.66~m, respectively. PW performs similarly to CW on LOS links but degrades more noticeably on NLOS links, indicating that received power captures only part of the path-dependent geometric reliability. \edit{The planar NLOS method in~\cite{8445980} and the 3D ZF--JP method in~\cite{Wei2011PIMRC} yield overall mean errors of 15.35~m and 11.25~m, respectively, with 30\% and 35\% of the links localized within 5~m. The 3D ZF--JP method, which uses both azimuth and elevation information, performs better than the planar method; nevertheless, both exhibit substantial degradation on NLOS links. This behavior is consistent with violations of their single-bounce propagation models by higher-order and poorly supported MPCs, for which the proposed LOS handling, point/line fallback, and measurement-derived reliability weighting provide greater robustness.}

\begin{figure}[t]
    \centering
    \includegraphics[width=0.8\columnwidth]{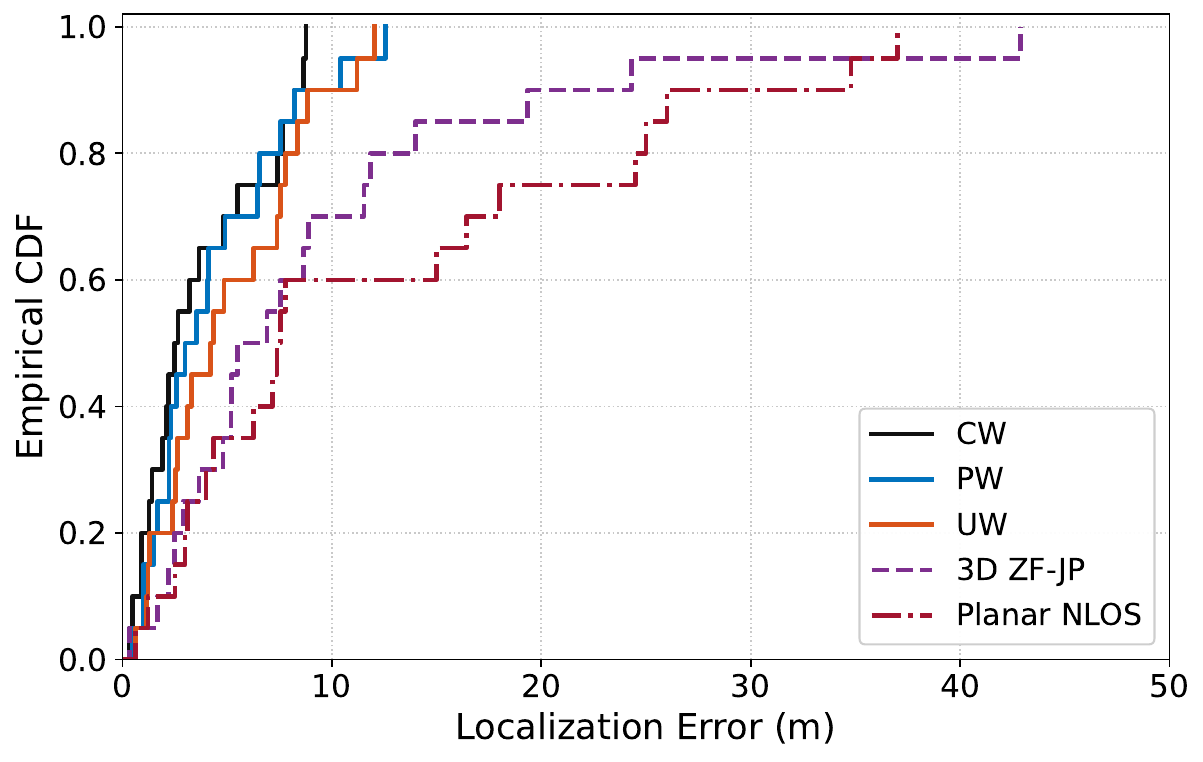}
    \caption{Empirical CDFs of the localization errors for the five methods using $K=5$ retained MPCs. Solid curves denote the three internal weighting variants, while dashed and dash-dotted curves denote the 3D ZF--JP and planar NLOS literature baselines, respectively.}
    \label{fig:cdf_compare_main}
\end{figure}

Fig.~\ref{fig:cdf_compare_main} compares the overall localization error distributions of the five methods. CW achieves the best CDF performance, followed by PW and UW, which use the same retained MPCs and point/line constraint construction. This ordering confirms that the proposed covariance-based weights provide more effective MPC reliability information than either received power or uniform weighting. \edit{The 3D ZF--JP and planar NLOS methods exhibit substantially heavier error tails, indicating reduced robustness to inconsistent or geometrically unfavorable MPCs in the measured indoor channels. Overall, the results demonstrate the benefit of combining measurement-derived uncertainty with the proposed point/line constraint fusion.}

\subsection{Sensitivity to the Number of MPCs}

\begin{figure}[t]
    \centering
    \includegraphics[width=0.8\columnwidth]{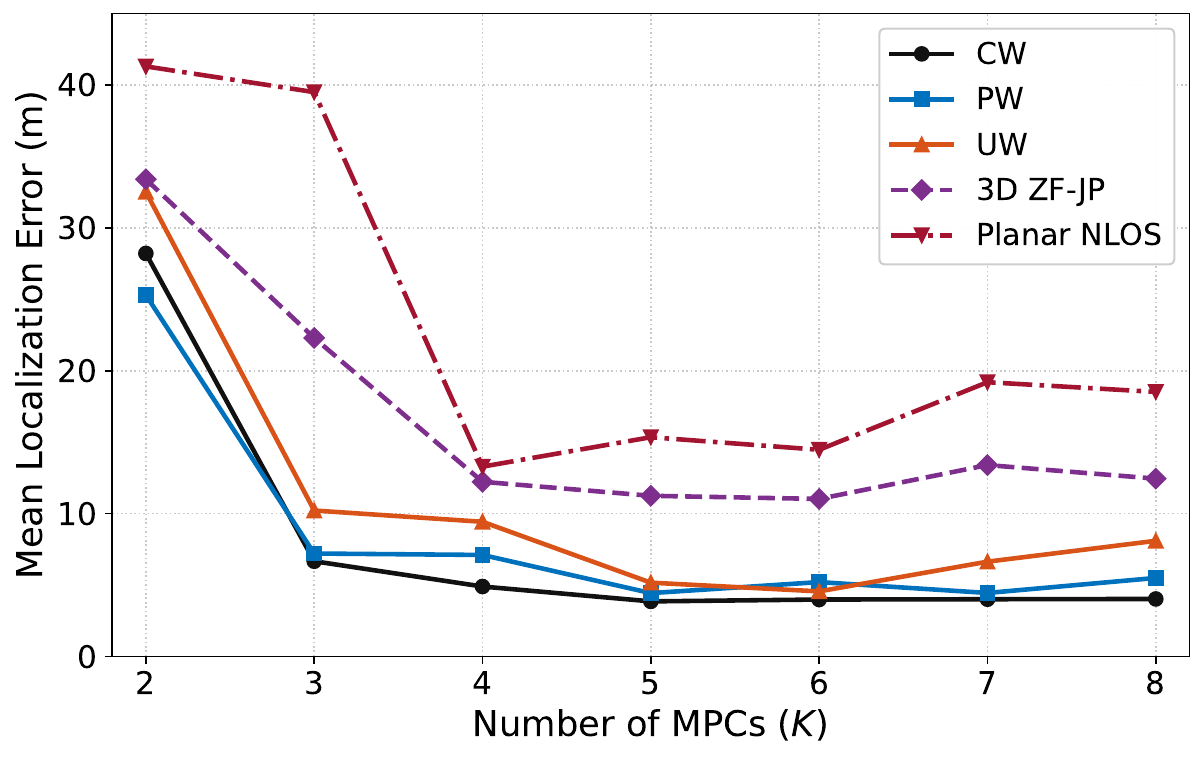}
    \caption{Overall mean localization error versus the number of retained MPCs $K$ for the five methods. Solid curves denote the three internal weighting variants, while dashed and dash-dotted curves denote the 3D ZF--JP and planar NLOS literature baselines, respectively.}
    \label{fig:mpc_num_sensitivity}
\end{figure}

Fig.~\ref{fig:mpc_num_sensitivity} shows the overall mean localization error versus the number of retained MPCs $K$ for the five methods. The comparison starts at $K=2$, the minimum path count for which all methods are defined. All methods exhibit relatively large errors at $K=2$, since a small number of inaccurate MPCs can produce an ill-conditioned estimate. The errors decrease substantially as additional MPCs are retained. Among the internal variants, CW reaches its lowest error at $K=5$ and remains stable at approximately 4~m for larger $K$, demonstrating robustness to MPC selection. PW generally outperforms UW but exhibits slightly larger fluctuations than CW. \edit{The 3D ZF--JP and planar NLOS baselines retain higher errors and larger fluctuations across $K$, reflecting their greater sensitivity to model-inconsistent MPCs. Retaining too many MPCs may also introduce late-arriving paths with low SNR or higher-order propagation, which can degrade methods that cannot sufficiently suppress unreliable observations. These results suggest that the preferred number of retained MPCs is environment-dependent and should be selected according to the available path quality and propagation conditions.}

\section{Conclusion}
\label{sec:conclusion}

A measurement-driven framework for uncertainty-aware single-anchor localization using 16.95~GHz indoor directional channel measurements was presented. Local directional neighborhoods were used to estimate per-MPC angular covariances, which were propagated into a type-aware fusion of LOS or feasible single-bounce point constraints and line fallbacks. Across 20 measured links, the proposed covariance-weighted method outperformed the power-weighted and unweighted variants as well as the 3D ZF--JP and planar NLOS baselines, demonstrating that local directional consistency provides reliability information beyond received power alone. The sensitivity results further showed that additional MPCs improve conditioning when few paths are available, whereas retaining too many may introduce unreliable late-arriving paths, motivating environment-dependent MPC selection. Future work will investigate adaptive MPC selection, cross-frequency generalization, and explicit delay-uncertainty modeling.

\bibliographystyle{IEEEtran}
\bibliography{Refs/refs}
\end{document}